\documentclass[11pt]{article}
\usepackage{amsfonts}
\usepackage{latexsym}
\usepackage{amssymb}
\usepackage{amscd}
\usepackage[leqno]{amsmath}

\newlength{\dinwidth}
\newlength{\dinmargin}
\def\hatdl{\widehat{d\l}}

\def\be{\begin{equation}}
\def\ee{\end{equation}}
\def\ben{\begin{displaymath}}
\def\een{\end{displaymath}}
\def\baa{\begin{eqnarray}}
\def\eaa{\end{eqnarray}}

\def\ba{\begin{array}}
\def\ea{\end{array}}

\def\Pcal{{\cal P}}

\makeatletter
\@addtoreset{equation}{section}
\makeatother

\def\phi{\varphi}

\def\a{\alpha}
\def\g{\gamma}

\def\b{\beta}

\def\l{\lambda}

\def\Th{\Theta}

\def\O{\Omega}

\def\Si{{\bf s}}

\def\sig{{\sigma}}

\def\cdiff{{\cal C}}

\def\Sfay{{S_{Fay}^{P_0}}}

\def\Dcal{{\cal D}}
\def\Acal{{\cal A}}
\def\Fcal{{\cal F}} 
\def\Ucal{{\cal U}}
\def\Wcal{{\cal W}}
\def\Lhat{{\hat{\L}}}

\def\pb{{\bf p}}
\def\qb{{\bf q}}
\def\rb{{\bf r}}
\def\sb{{\bf s}}

\def\B{{\bf B}}
\def\C{{\mathbb C}}

\def\CP1{{\mathbb C}P^1}

\def\la{\label}

\def\f{\frac}
\def\L{{\cal L}}
\def\p{\partial}

\def\tr{{\rm tr}}
\def\log{\ln}

\def\la{\label}

\def\f{\frac}
\def\L{{\cal L}}
\def\p{\partial}
\def\res{{\rm res}}

\def\det{{\rm det}}

\def\qb{{\bf q}}
\def\pm{\partial_m}

\def\pib{{\bf \pi}}

\newtheorem{remark}{Remark}
\newtheorem{theorem}{Theorem}
\newtheorem{corollary}{Corollary}
\newtheorem{lemma}{Lemma}

\begin{document}

\title{Isomonodromic tau-function of   Hurwitz Frobenius manifolds  and its applications}
%\shorttitle{}
\author{A. Kokotov and D. Korotkin}
%\Names{}
%\Email{korotkin@mathstat.concordia.ca, alexey@mathstat.concordia.ca}
\maketitle
\vskip0.5cm
{\bf Abstract.}
In this work we find the isomonodromic (Jimbo-Miwa)
tau-function corresponding to Frobenius manifold structures on  Hurwitz
spaces.
We discuss several  applications of this result.
First, we get 
an explicit expression for the  G-function (solution of Getzler's equation)
of the  Hurwitz Frobenius manifolds. 
Second, in terms of this tau-function we  compute the 
genus one correction to the free energy of hermitian two-matrix model. 
Third, we find the
Jimbo-Miwa tau-function of an arbitrary Riemann-Hilbert problem with quasi-permutation monodromy matrices.
Finally,  we get  a 
new expression (analog of genus one Ray-Singer formula) for the determinant of Laplace operator in the  Poincar\'e
metric on Riemann surfaces of an arbitrary genus.
\vskip0.5cm
Short title: ``Tau-function of Hurwitz Frobenius manifolds''
\vskip0.8cm
\section{Introduction}

The Hurwitz space $H_{g,N}$ is the space of equivalence classes of  
pairs ($\L$, $\pib$), where $\L$ is a compact Riemann surface of genus $g$ 
and $\pib$ is a 
meromorphic function of degree $N$. 
The Hurwitz space is stratified  according to multiplicities of poles 
and critical points
of function $\pi$ (see \cite{NatTur,Fulton});
in this paper we shall mainly work within the generic stratum 
$H_{g,N}(1,\dots,1)$, for which all critical points and poles of function $\pi$ 
are
simple. Denote the critical points of function $\pib$ by $P_1,\dots 
P_M$ ($M=2N+2g-2$ according to the Riemann-Hurwitz formula); the critical 
values
$\l_m=\pi(P_m)$ can be used as (local) coordinates on 
$H_{g,N}(1,\dots,1)$. The function $\pib$ defines  a realization of the 
Riemann surface $\L$ as an $N$-sheeted branched covering 
of $\CP1$ with ramification points $P_1,\dots, P_M$ and branch points 
$\l_m=\pib (P_m)$; enumerate the points at infinity
of the branched covering in some order and denote them by 
$\infty_1,\dots,\infty_N$. 
In a neighbourhood of the ramification point $P_m$ the  local 
coordinate
is chosen to be $x_m(P)=\sqrt{\pi(P)-\l_m}$, $m=1,\dots,M$; in a 
neighbourhood of any point $\infty_n$ the 
local parameter is $x_{M+n}(P)=1/\pi(P)$, $n=1,\dots,N$.

Fix a canonical basis of cycles $(a_\a,\,b_\a)$ on $\L$ and introduce 
the  prime-form $E(P,Q)$ on $\L$ and canonical meromorphic
bidifferential 
\be
W(P,Q)=d_P d_Q\log E(P,Q)
\la{bergdefin}\ee
 The bidifferential $W$ has the second order pole at $Q= P$ with the 
following
local behaviour: 
$$\f{W(P,Q)}{dx(P) dx(Q)}=\f{1}{(x(P)-x(Q))^2} 
+\f{1}{6}S_B(x(P))+o(1)\;,$$ 
where $x(P)$ is a local coordinate; $S_B(x(P))$ is the Bergman
projective connection. 

The central object of this paper is the function
$\tau(\l_1,\dots,\l_M)$ (the ``tau-function'') defined by the following system of equations:
\begin{equation} 
\f{\p}{\p \l_m}\log\tau = -\f{1}{12} S_B(x_m)|_{x_m=0}\;,\hskip0.8cm 
m=1,\dots,M\; ;
\label{deftau}
\end{equation}
compatibility of this system can  be obtained as  a simple corollary of the Rauch 
variational formulas \cite{KokKor1}.
In global terms, $\tau$ is a horizontal holomorphic section of the flat 
holomorphic line bundle ${\cal T}_B$ (see \cite{KokKor1})
over the space $\widehat{H_{g,N}}(1,\dots,1)$, which covers  
$H_{g,N}(1,\dots,1)$, and consists of pairs
(weakly marked Riemann surface $\L$;  meromorphic function $\pi$ with 
simple poles and critical points).
This covering space appears due to dependence of the 
bidifferential $W$ (and, therefore, the Bergman projective connection) 
on the choice of homology basis on $\L$.

In the Frobenius manifolds theory \cite{D}, apart from the prepotential (solutions of WDVV
equations), an important role is played by the so-called $G$-function, which is 
the genus one  free energy, corresponding to
a given Frobenius manifold (the prepotential itself equals to the planar limit of the free
energy). It was conjectured by Givental \cite{Givental} and proved by Dubrovin-Zhang \cite{DZ1} that 
the $G$-function can be expressed in terms of Jimbo-Miwa tau-function of the isomonodromic problem
corresponding to a given Frobenius manifold. 

In \cite{D}   Frobenius manifold structures were found on an arbitrary Hurwitz space;
so far this is, probably, one of most well-understood classes of Frobenius manifolds
(alternative structures of Frobenius manifolds on Hurwitz spaces were recently found in 
\cite{Vaska1,Vaska2}).
As it was recently proved in \cite{KokKor2}, the definition of  
isomonodromic tau-function $\tau(\l_1,\dots,\l_M)$ of Hurwitz Frobenius manifolds
from \cite{D} is equivalent to  (\ref{deftau}).

The same tau-function (\ref{deftau}) appears as one of two multipliers 
in the   Jimbo-Miwa tau-function 
corresponding to another class of Riemann-Hilbert problems - the Riemann-Hilbert problems  with quasi-permutation 
monodromy matrices \cite{Kor03}.

In \cite{KokKor1} it was also revealed the role of the 
function $\tau$ in the problem of holomorphic factorization of
the determinant of the Laplacian  on Riemann surfaces: namely, up to a 
factor involving an appropriate regularized Dirichlet integral and 
the matrix 
of  $b$-periods of
a Riemann surface, the determinant of Laplace operator (in the 
Poincar\'e metric) acting in the trivial line bundle over Riemann surface is 
given by $|\tau|^2$.

Another important area where the same tau-function appeared recently is the
large $N$ limit of Hermitian two-matrix model \cite{EKK}; in this paper it was realized that the
subleading correction to the free energy of such models formally almost
 coincides with the $G$-function of Hurwitz Frobenius manifolds. In particular, the isomonodromic tau-function  (\ref{deftau}) is the most non-trivial ingredient of this sub-leading correction.

For $N=2$ and arbitrary $g$ the  Riemann surface $\L$ is hyperelliptic, 
and can be defined by equation 
$w^2=\prod_{m=1}^{2g+2}(\l-\l_m)$.
In this case $\tau=\det {\bf A} \prod_{m\neq n}^M(\l_m-\l_n)^{1/4}$ 
\cite{KitKor}, where ${\bf A}$ is the matrix of $a$-periods 
of non-normalized holomorphic 
differentials $\l^{\a-1}d\l/w\;,\; \a=1,\dots,g$. 
In other simple case, when $g=0,1$ and $N$ is arbitrary,the function 
$\tau$ was found in \cite{KokKor1}; 
this result allowed to compute  the $G$-function of  Frobenius manifold 
related to the  extended affine Weyl group 
$\tilde W(A_{N-1})$ (originally found in \cite{Strachan})
 and the $G$-function of Frobenius manifold related to  the Jacobi 
group $J(A_{N-1})$ (conjectured in \cite{Strachan}).

The goal of this paper is to compute the   tau-function 
of the isomonodromy problem  corresponding to Frobenius structures on an arbitrary Hurwitz space 
$H_{g,N}(1,\dots,1)$. 

Consider the divisor $\Dcal$ of the differential $d\pib$:
$\Dcal=\sum_{k=1}^{M+N} d_k D_k$,
where $D_m=P_m\;,\;\; d_m=1$ for $m=1,\dots,M$ and 
$D_{M+n}=\infty_{n}\;,\;\; d_{M+n}=-2$ for $n=1,\dots,N$.
 Here and below, if the argument of a differential coincides with a 
point $D_k$ of divisor $\Dcal$, we evaluate this
differential at this point with respect to local parameter $x_k$.
In particular, for the prime form we shall use the following 
conventions:
\begin{equation}
E(D_k,D_l):= {E(P,Q)}\sqrt{dx_k(P)}\sqrt{dx_l(Q)}|_{P=D_k,\;Q=D_l}\;,
\end{equation}
for $k,l=1,\dots,M+N$. 
The next notation corresponds to prime-forms, evaluated at points of 
divisor $\Dcal$  with respect to  only one argument:
\begin{equation}
E(P,D_k):= {E(P,Q)}\sqrt{dx_k(Q)}|_{Q=D_k}\;,
\end{equation}
$k=1,\dots,M+N$;
in contrast to $E(D_k,D_l)$, which are just scalars, $E(P,D_k)$ are 
$-1/2$-forms with respect to $P$.

 Denote by $v_1,\dots,v_g$  the normalized ($\oint_{a_\alpha} 
v_\beta=\delta_{\a\b}$ ) holomorphic differentials on $\L$;
$\B_{\a\b}=\oint_{b_\alpha} v_\beta$ is the corresponding matrix of 
$b$-periods; $\Th(z|\B)$ is the theta-function.
Let us dissect the Riemann surface $\L$ along its basic cycles to get 
its fundamental polygon $\Lhat$; 
choose some initial point 
$P\in\Lhat$ and introduce the corresponding vector of Riemann constants 
\begin{equation}\label{KP}
K^P_\a=\frac{1}{2}+\frac{1}{2}\B_{\a\a}-\sum_{\b\neq
\a}\oint_{a_\b}(v_\b(Q)\int_P^Q v_\a)\ ;\ \ \a=1, \dots, g
\end{equation}
and the Abel map $[\Acal_P]_\a(Q)=\int_P^Q v_\a$,
computed along path which does not intersect $\p\Lhat$.  

The following theorem, together with its applications, is the main 
result of this paper

\begin{theorem}
Assume that the fundamental domain $\Lhat$ is chosen in such a way that
\be
\Acal(\Dcal)+2K^P=0.
\la{defind}
\ee 
The isomonodromic tau-function (\ref{deftau}) of a Frobenius manifold associated to the Hurwitz space  
$H_{g,N}(1,\dots,1)$ is given by the following expression:
\begin{equation}
\tau = {\Fcal}^{2/3}  \prod_{k,l=1\;\;k< l}^{M+N} [E(D_k, 
D_l)]^{\frac{d_k d_l}{6}} \; 
\la{tauint}
\end{equation}
where the quantity $\Fcal$ defined by
\begin{equation}
\Fcal = \frac{[d\pib(P)]^{\f{g-1}{2}}}{\Wcal (P)}
\left\{\prod_{k=1}^{M+N}[ E (P,D_k)]^{\f{(1-g)d_k}{2}}\right\}
{\sum_{\a_1,\dots,\a_g=1}^g \frac{\partial^g\Theta(K^P)}{\p 
z_{\a_1}\dots \p z_{\a_g}} v_{\a_1}(P)\dots v_{\a_g}(P)}
\la{Fdef}
\end{equation}
is independent of the point $P\in \L$. Here
 $\Theta$ is the theta-function of $\L$;
integer vector $\rb$ is defined by (\ref{defind}); 
$\langle\;,\;\rangle$ is the scalar product in ${\mathbb C}^g$,
 ($\langle x,y\rangle=\sum_{\a=1}^gx_\a y_\a$);
\be
\Wcal(P):= {\rm \det}_{1\leq \a, \b\leq g}||v_\b^{(\a-1)}(P)||
\la{Wronks}
\ee
denotes the Wronskian determinant of holomorphic differentials
 at the point $P$.
\end{theorem}

The proof of this theorem is contained in Section 2.
% and is rather 
%indirect; the first main ingredient of the proof is a variational formula 
%for
%the part of expression $\Fcal$  which depends only on the moduli of 
%Riemann surface $\L$ (\cite{Fay92}, p.58);
%we notice that $\Fcal$ is a natural generalisation of genus 1 
%expression $\Theta^{{\prime}}(\f{\B+1}{2})$ 
%(which is equal to $\eta^3(\B)e^{-\pi\B/4}$, where $\eta$ is the  
%Dedekind eta-function)
% to higher genus.

% The second main ingredient of the proof is the technique of variation 
%and holomorphic factorization of a Dirichlet integral
%on $\L$. 

%Without any modification the same formula gives the Bergman 
%tau-function on an arbitrary stratum of Hurwitz space
%(when divisor $\Dcal$ of differential $d\pi$ has an arbitrary allowed 
%(i.e. such that  $\sum d_k =2g-2$) set of multiplicities $d_k$).

In section 3 we discuss applications of the formula (\ref{tauint}).
First, in Sect.3.1 we show  how  to find the
 $G$-function of Frobenius manifolds from \cite{D}
corresponding to Hurwitz spaces; the resulting formula looks as follows:
\be
G=-\frac{1}{2}\log\tau-\frac{1}{48}\sum_{m=1}^M\log\,{\rm 
Res}\,_{P_m}\frac{\phi^2}{d\l}\;,
\la{Gint}
\ee
where $\phi$ is a primary differential defining the Frobenius manifold \footnote{The tau-function 
$\tau$ is defined according to original formula of Jimbo-Miwa \cite{JimMiw}; the isomonodromic
tau-function $\tau_I$  defined in \cite{DZ1} is related to $\tau$ as
follows: $\tau_I=\tau^{-1/2}$ (we thank V.Shramchenko for this observation).
Here we prefer the convention of \cite{JimMiw}, since it is this definition which guarantees the holomorphy of
the tau-function in general case.}.
In Section 3.2 we  use the tau-function (\ref{tauint}) to compute the genus one correction to
the free energy of hermitian two-matrix matrix model.
In section 3.3 the  formula (\ref{tauint}) is used to get a new expression 
(valid up to a constant independent of moduli of the Riemann surface)
for  the determinant of the Laplacian on Riemann
surface $\L$ in Poincar\'e metric. A formula for ${\rm det}\Delta$ in
Arakelov metric was proved by Fay \cite{Fay92}; combining this formula with  Polyakov's formula
relating determinants of Laplacians in different conformal metrics on the same Riemann surface, one can
get an expression for ${\rm det}\Delta$ in the Poincar\'e metric. The expression we derive here is
different, and  is given by the modulus square of the tau-function (\ref{tauint}) multiplied by the
exponent of an appropriate Dirichlet integral.
In section 3.4  we show how to apply the
formula (\ref{tauint}) to find the Jimbo-Miwa tau-function 
of another class of Riemann-Hilbert problems -  the ones with arbitrary
quasi-permutation monodromy matrices \cite{Kor03}. 

This paper is based on the authors' preprint \cite{KKBonn}.

\section{Proof of the main theorem}

\subsection{Variational formulas on the spaces of branched coverings}
Here we establish the formulae describing the variations of basic holomorphic objects
(holomorphic differentials, the canonical bidifferential, the prime-form, the vector of Riemann constants, etc)
on the Riemann surface $\L$ under the variation of a critical value of the map $\pi\ :\  \L\rightarrow {\mathbb C}P^1$.

With a slight abuse of 
terminology we denote the branched covering
of the Riemann sphere defined by the function $\pi$ on the Riemann 
surface $\L$ by the same letter $\L$; the 
coordinate on the covered Riemann sphere will be denoted by $\l$.
The zeros of $d\pi$ are the ramification point of the branched covering 
$\L$; the local parameter in a neighbourhood
of $P_m$ is $x_m=\sqrt{\l-\l_m}$, according to the notations from 
introduction.

First, we recall the properties of the  prime form $E(P, Q)$ (see \cite{Fay73,Fay92}), which 
 is an antisymmetric  $-1/2$-differential with respect to both $P$ and $Q$:

\begin{itemize}
\item Under tracing of $Q$ along the cycle $a_\a$ the prime-form remains invariant; under 
the tracing along  $b_\alpha$ it gains the factor
\begin{equation}\label{primetwist}
\exp(-\pi i \B_{\a\a}-2\pi i\int_P^Q v_\a)\;.
\end{equation}  
\item The prime-form can be expressed  in terms of the canonical meromorphic  bidifferential
$W(P, Q)$ as follows:
\begin{equation}\label{primrep}
E^2(P, Q)dx(P)dy(Q)=\lim_{P_0\to P, Q_0\to
Q}(x(P_0)-x(P))(y(Q)-y(Q_0))\exp\left(-\int_{P_0}^{Q_0}\int_P^Q W(\, \cdot\,, \,\cdot\,)\right),
\end{equation}
\item At the diagonal $Q= P$  the prime-form has the first order zero; the following asymptotics holds:
$$E(x(P), x(Q))\sqrt{dx(P)}\sqrt{dx(Q)}=$$
\begin{equation}\label{primas}
(x(Q)-x(P))\left(1-\frac{1}{12}S_B(x(P))(x(Q)-x(P))^2+O((x(Q)-x(P))^3\right),
\end{equation}
as $Q\to P$,
where $S_B$ is the Bergman projective connection and $x(P)$ is an arbitrary local parameter.
\end{itemize}

Now for any two points $P,Q\in\Lhat$ we define
\be
\Si(P,Q):=\exp\left\{-\sum_{\a=1}^g\oint_{a_\a}v_\a(R)\log \frac{E(R, 
P)}{E(R, Q)}\right\}
\label{sigma}
\ee
This object is a (multi-valued) non-vanishing holomorphic  $g/2$-differential  on 
$\Lhat$ with respect to $P$ and  non-vanishing holomorphic 
$-g/2$-differential with respect to $Q$. 
 Under tracing along the cycles $a_\a$ and $b_\a$
 it gains 
the multipliers $1$ and
 $\exp[(g-1)\pi i {\bf B}_{\a\a}+2\pi i K^P_\a]$  with respect to $P$ 
and the multipliers $1$ and 
$\exp[(1-g)\pi i {\bf B}_{\a\a}-2\pi i K^Q_\a]$ with respect to $Q$.

As in the case of the prime form, if one  of the arguments coincides with a 
point of $\Dcal$, we evaluate $\Si$ in 
the corresponding local parameter: 
\be
\Si(D_k,Q):= \Si(P,Q)(d x_k(P))^{-g/2}|_{P=D_k}
\ee
where $k=1,\dots,M+N$.
For arbitrary two points $P,Q\not\in D$ we introduce the following 
notation:
\be
\sig(P,Q):= \Si(P,Q)(d\pi(P))^{-g/2}(d\pi(Q))^{g/2}
\ee

Let us fix two points $P_0,\,Q_0\in\L$ (for convenience in the sequel 
we assume that these points do not 
coincide with points of $\Dcal$), 
and introduce another object which plays an important role below,
the following (non-single-valued) holomorphic 1-differential on $\Lhat$:
\begin{equation}\label{Fdiff}
\omega(P)=\Si^2(P, Q_0)E(P, 
P_0)^{2g-2}(d\pi(Q_0))^{g}(d\pi(P_0))^{g-1}\;;
\end{equation}
in agreement with the previous notations, $\omega(D_k):=\f{\omega(P)}{d 
x_k(P)}|_{P=D_k}$.
The differential $\omega(P)$  has multipliers $1$ and $\exp(4\pi i 
K_\a^{P_0})$ along the basic cycles $a_\a$ and $b_\a$ 
respectively.  
The only zero of the $1$-form $\omega$ on $\Lhat$ is $P_0$; its 
multiplicity equals $2g-2$.

Consider the following Schwarzian derivative 
(which depends on the chosen point $P_0$, but is obviously independent 
of the point $Q_0$ from (\ref{Fdiff}))
\be
\Sfay(x(P)):=\left\{\int^P \omega(P),\; x(P)\right\}\;,
\la{Sfa}
\ee
where $x(P)$ is a local coordinate on $\L$; $\Sfay$ is a projective 
connection (see e.g. \cite{Tyurin}) on $\L$;
this object was introduced and exploited by Fay \cite{Fay92} in a 
different form (and without mentioning that
it is a projective connection).

 Introduce also the following holomorphic non-single-valued $g(1-g)/2$-differential on $\Lhat$ which 
has multipliers
$1$ and $\exp\{-\pi i (g-1)^2\B_{\a\a}-2\pi i (g-1) K_\a^P\}$ along 
basic
cycles $a_\a$ and $b_\a$, respectively:
\begin{equation}\label{c}
\cdiff(P)=\frac{1}{\Wcal[v_1, \dots, v_g](P)}\sum_{\a_1, \dots, \a_g=1}^g
\frac{\partial^g\Theta(K^P)}{\p z_{\a_1}\dots \p z_{\a_g}}
v_{\a_1}\dots v_{\a_g}(P)\;,
\end{equation}
where $\Wcal(P)$ is the  Wronskian (\ref{Wronks}) from the introduction.
 
%\be
%W(P):= {\rm \det}_{1\leq \a, \b\leq 
%g}||\f{1}{(\a-1)!}v_\b^{(\a-1)}(P)||\;.
%\la{Wronks}
%\ee
The differential $\cdiff$ is an essential ingredient of the Mumford measure 
on the moduli space of Riemann surfaces of given genus
\cite{Fay92}. For $g>1$ the multiplicative differential $\Si$ (\ref{sigma}) is 
expressed in terms of $\cdiff$ as follows \cite{Fay92}:
\be
\Si(P,Q)= \left(\f{\cdiff(P)}{\cdiff(Q)}\right)^{1/(1-g)}
\la{Csi}
\ee
According to Corollary 1.4 from\cite{Fay92}, $\cdiff(P)$ does not have any zeros.
Moreover, this object admits the following alternative representation: 
\begin{equation}\label{repC}
\cdiff(P)=\frac{\Theta(\sum_{\a=1}^{g-1}\Acal_P(R_\a)+\Acal_Q(R_g)+K^P)\prod_{\a<\b}E(R_\a, R_\b)\prod_{\a=1}^g\Si(R_\a, P)
}
{\prod_{\a=1}^g E(Q, R_\a)\,{\rm det}\,||v_\a(R_\b)||_{\a, \b=1}^g\Si(Q, P)}.
\end{equation}
where $Q, R_1, \dots, R_g\in \L$ are arbitrary points of ${\cal L}$.

The following Theorem describes the behaviour of the basic holomorphic differentials $v_\a$, the matrix $\B$ of $b$-periods, the canonical bidifferential $W(P, Q)$, the prime form $E(P, Q)$, the vector of Riemann constants $K^P$, and the multiplicative differentials $\Si(P, Q)$ and $\cdiff(P)$ under variations of the critical values $\l_m$.

From now on we use the notation
$$\pm T(P_m):=\frac{d T(x_m)}{dx_m}\Big|_{x_m=0}$$
for the derivative of a tensor $T(x_m)(dx_m)^r$ of a (possibly fractional) weight $r$ at the critical point $P_m$ calculated with respect to the local parameter $x_m$. 

\begin{theorem}\label{mainvar} Let the coordinates $\l(P)=\pi(P)$ and $\l(Q)=\pi(Q)$ of the points $P$ and $Q$ do not change when the covering
$\pi:\L\rightarrow {\mathbb C}P^1$ deforms. Under the convention that all the tensor objects with arguments $P$, $Q$ and $Q_0$ are calculated in the local parameter $\l$ lifted from the base ${\mathbb C}P^1$ of the covering $\pi$ and all the tensor objects with argument $P_m$ are calculated in the local parameter $x_m=\sqrt{\l-\l_m}$,  the following variational formulae hold

\begin{equation}\label{differential}
\frac{\partial v_\a(P)}{\partial \l_m}=\frac{1}{2}W(P, P_m)v_\a(P_m),
\end{equation}

\begin{equation}\label{bperiods}
\frac{\partial \B_{\a\b}}{\partial \l_m}=\pi i v_\a(P_m)v_\b(P_m),
\end{equation}

\begin{equation}\label{bergmanbidiff}
\frac{\partial W(P, Q)}{\partial \l_m}=\frac{1}{2}W(P, P_m)W(P_m, Q),
\end{equation}

\begin{equation}\label{primeform}
\frac{\partial E(P, Q)}{\partial \l_m}=-\frac{1}{4}
\left[\pm\log\frac{E(P, P_m)}{E(Q, P_m)}\right]^2,
\end{equation}

\begin{equation}\label{Riemconst}
\frac{\partial K^P_\a}{\partial
\l_m}=\frac{1}{2}v_\a(P_k)\pm\log\left(\Si(P_m, Q_0)E(P_m,P)^{g-1}\right)
-\frac{1}{4}\pm v_\a(P_m)\;,
\end{equation}

$$
\frac{\partial \sigma(P, Q)}{\partial \l_m}=
\frac{1}{4}
\left\{\pm\log\frac{E(P_m, P)}{E(P_m, Q)}\right\}\left\{\pm\log\left[
\Si(P_m, Q_0)^2 E(P_m, P)^{g-1}E(P_m, Q)^{g-1}\right]\right\}
$$
\begin{equation}\label{sigmadiff}
-\frac{1}{4}\pm^2\log\frac{E(P_m, P)}{E(P_m, Q)}\;,
\end{equation}

\begin{equation}\label{cmultdiff}
\frac{\partial \cdiff(P)}{\partial \l_m}=-\frac{1}{8}(S_B-S^P_{Fay})(P_m)\;,
\end{equation}
where the expressions in the right hand sides of (\ref{Riemconst}), (\ref{sigmadiff}) and (\ref{cmultdiff}) are $Q_0$-independent. 
\end{theorem}

\begin{remark} \rm
Formally,  the expressions (\ref{differential}-\ref{cmultdiff}) are complete analogs of 
variational formulas (3.21), (3.22), (3.24) and (3.25) from \cite{Fay92}. 
However,  Th. \ref{mainvar} cannot be obtained as a straightforward consequence of the 
formulae from \cite{Fay92}. In the scheme of moduli deformation used in \cite{Fay92} the $C^\infty$-surface $\L$ is fixed and the systems of coordinates defining the complex structures on $\L$ vary, while we 
use the branch points as the local coordinates on the moduli space. Although it is not too difficult
to establish a direct correspondence between the two deformation schemes for the objects which don't
depend on a point of the Riemann surface (see for example \cite{KokKor1}), for the point-dependent 
objects (like all the  objects listed in the theorem except the matrix of $b$-periods) it is much less trivial, since the fixing of the argument in the 
two schemes is essentially different. Therefore, we prove the theorem independently; formally
the proof looks very similar to  \cite{Fay92}.
\end{remark}
%\begin{remark} {\rm A physicist may omit the foregoing proof. The simple substitution of the "Beltrami differential" 
%$$\mu=-\frac{\pi}{2}\delta(\, \cdot -P_m),$$
%which "describes" the variation of the complex structure of the Riemann surface $\L$ under the variation of the critical value $\l_k$ in Fay's formulae (3.21)-(3.25), should be convincing. A closely related trick (with  true $L^\infty$ Beltrami differentials) was justified and applied to 
% variational formulas for the determinants of the laplacians  in  (\cite{KokKor1}, p. 53-54). 
%} 
%\end{remark}

{\bf Proof.} An elementary proof of formulae (\ref{differential}-\ref{bergmanbidiff}) can be found in 
\cite{KokKor1}. As in \cite{Fay92}, formula (\ref{primeform}) immediately follows from (\ref{bergmanbidiff}) and (\ref{primrep}). Let us prove (\ref{Riemconst}).

One may assume that the projections of $a$- and $b$-cycles on $\l$-plane
 do not move when the covering deforms.
Varying the right hand side of (\ref{KP}) via (\ref{bperiods}) and ({\ref{differential}) and taking into account (\ref{bergdefin}), we get
$$\partial_{\l_m}K^{P}_\a=\frac{\pi i}{2}v_\a(P_m)^2-\sum_{\b\neq \a}\oint_{a_\b}
\frac{1}{2}\left\{\partial_\l\pm\log E(\l, P_m)v_\b(P_m)\int_P^\l
v_\a\right\}d\l
$$ 
$$- \oint_{a_\b} v_\b(\l)\int_P^\l \left(
\frac{1}{2}\partial_{\l'}\pm\log E(\l', P_m)v_\a(P_m)\right)d\l'
=
$$
$$
\frac{\pi i}{2}v_\a(P_m)^2-\frac{1}{2}\sum_{\b\neq\a}v_\b(P_m)\oint_{a_\b}\left\{(\partial_\l\pm\log
E(\l, P_m))\int_P^\l v_\a\right\}d\l
$$
$$-\frac{v_\a(P_m)}{2}\sum_{\b\neq \a}\oint_{a_\b}
v_\b(\l)\pm\log\frac{E(\l, P_m)}{E(P, P_m)}=
$$
$$\frac{\pi i}{2}v_\a(P_m)^2+\frac{g-1}{2}v_\a(P_m)\pm\log E(P,
P_m)-\frac{1}{2}\sum_{\b\neq \a}v_\b(P_m)
\oint_{a_\b}\left\{ \partial_\l\pm E(\l, P_m)\int_P^\l v_\a
\right\}d\l
$$
$$
-\frac{v_\a(P_m)}{2}\sum_{\b\neq \a}\oint_{a_\b} v_\b(\l)\pm\log E(\l,
 P_m)=
$$
$$=\frac{\pi i}{2}v_\a(P_m)^2+\frac{v_\a(P_m)}{2}\pm\log E^{g-1}(P,
P_m)-\frac{1}{2}\sum_{\b\neq \a}v_\b(P_m)\oint_{a_\b}\left\{
\partial_\l\pm E(\l, P_m)\int_P^\l v_\a\right\}d\l$$
$$+\frac{v_\a(P_m)}{2}\pm\log \Si(P_m, Q_0)+\frac{v_\a(P_m)}{2}\oint_{a_\a}\,v_\a(\l)\pm\log E(P_m, \l)=$$
\begin{equation}
\label{vykladka}
\frac{v_\a(P_m)}{2}\pm \log \Si(P_m, Q_0)E^{g-1}(P, P_m)
+\frac{\pi i}{2}v_\a(P_m)^2
+\frac{1}{2}\sum_{\b=1}^gv_\b(P_m)\oint_{a_\b}\left\{ \pm\log E(\l, P_m)\right\}
v_\a(\l)\, ,
\end{equation}
where the last equality is obtained via integration by parts (which is
possible since the prime form has no 
twists along the $a$-cycles). Following Fay (\cite{Fay92}), we notice
that, due to (\ref{primetwist}), the sum 
of the last two terms in the latter expression  coincides with the
following integral over the boundary of the 
fundamental polygon $\hat\L$
\begin{equation}\label{integralF}-\frac{1}{8\pi i}\oint_{\partial
\hat\L} v_\a(\l)\left(\pm\log E(\l, P_m)
\right)^2\,.\end{equation}   
From asymptotics (\ref{primas}) and the Cauchy formula it follows that integral (\ref{integralF}) coincides with
$$-\frac{1}{4}v_\a'(x_m)\Big|_{x_m=0}\equiv -\frac{1}{4}\pm v_\a (P_m) \, ,$$ which gives equation (\ref{Riemconst}).

Let us prove (\ref{sigmadiff}). Due to (\ref{differential}) and (\ref{primeform}), we have
$$\partial_{\lambda_m}\log 
\Si(P,Q)=-\partial_{\l_m}\sum_{\beta=1}^g\oint_{a_\b}v_\beta(\l)
\log\frac{E(\l, P)}{E(\l, Q)}\,=
-\frac{1}{2}\sum_{\beta=1}^g\oint_{a_\b}[\partial_\l\pm\log E(P_m, \l)]v_\beta(P_m)\log\frac{E(\l, P)}{E(\l, Q)}\,d\l$$
\begin{equation}\label{vvss1}
+\frac{1}{4}\sum_{\beta=1}^g\oint_{a_\b}v_\b(\l)\left\{\left(\pm\log\frac{E(\l, P_m)}{E(P, P_m)}\right)^2-
\left(\pm\log\frac{E(\l, P_m)}{E(Q, P_m)}\right)^2
\right\}\,:=\Sigma_1+\Sigma_2.
\end{equation}

To simplify the first sum in (\ref{vvss1}) we integrate it by parts,  rewrite the resulting expression as the integral over the boundary of the fundamental polygon and apply the Cauchy theorem:
$$\Sigma_1=\frac{1}{2}\sum_{\beta=1}^g\oint_{a_\b}v_{\beta}(P_m)\pm\log
E(P_m, \l)\partial_{\l}\log\frac{E(\l, P)}{E(\l,
Q)}\,d\l=-\frac{1}{8\pi i}\oint_{\partial \hat{\L}}(\pm\log E(P_m,
\l))^2\partial_\l
\log\frac{E(\l, P)}{E(\l, Q)}\,d\l=$$
\begin{equation}\label{primo}
-\frac{1}{4}\left[\pm^2\log\frac{E(P_m, P)}{E(P_m, Q)}+(\pm\log E(P, P_m))^2-(\pm\log E(Q, P_m))^2\right].
\end{equation}
Here we used the fact that the function $$\l\mapsto \partial_\l\log\frac{E(\l, P)}{E(\l, Q)}$$  is single-valued on $\L$ 
and that
$$\oint_{a_\b}\partial_\l\log\frac{E(\l, P)}{E(\l, Q)}\,d\l=0,$$
due to the single-valuedness of the prime form along the $a$-cycles.

The second sum in (\ref{vvss1}) can be rewritten as
$$\Sigma_2=\frac{1}{4}\sum_{\beta=1}^g\oint_{a_\beta}v_\beta(\l)\left\{
2\pm \log E(\l, P_m)\pm\log\frac{E(Q, P_m)}{E(P, P_m)}+\pm\log(E(P,
P_m)E(Q, P_m))\pm\log
\frac{E(P, P_m)}{E(Q, P_m)}
\right\}=$$
$$
-\frac{1}{2}\pm\log\frac{E(P, P_m)}{E(Q, P_m)}\sum_{\beta=1}^g\left\{\pm\oint_{a_\b}v_\b(\l)\log\frac{E(\l, P_m)}{E(\l, Q_0)}
\right\}+\frac{g}{4}\pm\log(E(P, P_m)E(Q, P_m))\pm\log\frac{E(P, P_m)}{E(Q, P_m)}$$
\begin{equation}\label{secundo}
=\frac{1}{4}\pm\log\frac{E(P, P_m)}{E(Q, P_m)}\pm\log\left[\Si^2(P_m, Q_0)E^g(P, P_m)E^g(Q, P_m)\right].
\end{equation} 
Relation (\ref{sigmadiff}) follows from (\ref{primo}) and (\ref{secundo}).

Now we are in a position to prove the main statement (\ref{cmultdiff}) of the theorem.

Let us first rewrite the definition of Fay's projective connection
(\ref{Sfa}) in a local parameter $\zeta$ as follows:
\begin{equation}\label{proektivnaja}
S^P_{Fay}(\zeta)=2\partial_{\zeta \zeta}^2\log[\Si(\zeta, Q_0)E(\zeta, P)^{g-1}(d\zeta)^{-1/2}]-2(\partial_\zeta\log[\Si(\zeta, Q_0)E(\zeta, P)^{g-1}(d\zeta)^{-1/2}])^2\,.
\end{equation}

Similarly to (\cite{Fay92}), to prove (\ref{cmultdiff}) we are to vary the logarithm of the right
hand side of 
expression (\ref{repC}) and pass to the limit $R_1, \dots, R_g\to P$, and then $Q\to P_m$.

In what follows all the tensor objects with arguments $P, Q, R_1, \dots, R_g$ are calculated in the local parameter $\l$ and, as before, the appearance of the argument $P_m$  means means that the corresponding tensor is calculated in the local parameter $x_m$ at the point $x_m=0$.    

The next lemma describes the variation of the determinant ${\rm det}\,||v_\a(R_\b)||$ from the denominator of expression (\ref{repC}).
\begin{lemma}\label{denominator} Assume that none of the points
$R_1,\dots R_g$  coincide with the ramification points $\{P_m\}$, and
the  projections of the points $\{R_\a\}$ on $\l$-plane don't depend on
$\{\l_m\}$. Then the following variational formula holds
\begin{equation}\label{denden}
\lim _{R_1, \dots, R_g\to P}\frac{\partial \log {\rm
det}\,||v_\a(R_\b)||}{\partial \l_m}
=-\frac{1}{2}\sum_{\a, \b=1}^g
\partial^2_{z_\a z_\b}\log \Theta(K^P-{\cal A}_P(P_m))v_\a(P_m)v_\b(P_m).
\end{equation}
\end{lemma}
This lemma is an immediate corollary of   (\ref{differential}) and the
formula (35) from \cite{Fay73}, which expresses the second
derivative of the theta-function $\Theta (\Acal|_P^Q-K)$ in terms of
the bidifferential $W$. $\Box$

Using (\ref{bperiods}), we can represent  the variation of the theta-functional term from the numerator of (\ref{repC}) as follows 
$$\partial_{\l_m}\log \Theta (\sum_{\g=1}^{g-1}\Acal_P(R_\g)+\Acal_Q(R_g)+K^P\, | \,{\B})=$$
\begin{equation}\label{summadvuh}
\sum_{\g=1}^g
\left[\partial_{\l_m}\int_{Q+(g-1)P}^{\sum_{\g=1}^g R_\g}v_\a+\partial_{\l_m}K^P_\a
\right]\frac{\partial \log\Theta}{\partial z_\a}+\pi i\sum_{\a, \b=1}^g \frac{\partial \log \Theta}{\partial \B_{\a\b}}v_\a(P_m)v_\b(P_m).
\end{equation}
We have 
$$\partial_{\l_m}\int_{Q+(g-1)P}^{\sum_{\g=1}^g
R_\g}v_\a=\frac{1}{2}\int_{Q+(g-1)P}^{\sum_{\g=1}^g R_\g}\pm\partial_\l\log E(\l, P_m)v_\a(P_m)\,d\l=$$
\begin{equation}\label{vintegral} 
=\frac{1}{2}\pm\log E(P, P_m) v_\a(P_m)-\frac{1}{2}\pm\log E(Q, P_m) v_\a(P_m)+o(1) \end{equation}  
as $R_1, \dots, R_g \to P$.
Now from (\ref{summadvuh}), (\ref{vintegral}), (\ref{Riemconst}), the heat equation for the theta-function
and the obvious relation
$$
\pm\log\Theta(K^P-\Acal_P(P_m))\equiv\partial_{x_m}\log\Theta(\int_{x_m}^P\vec{v}+K^P)|_{x_m=0}
=-\sum_{\a=1}^g(\log \Theta)_{z_\a}v_\a(P_m)
$$
 it follows that
$$\lim_{R_1, \dots, R_g\to P} \partial_{\l_m}\log \Theta (\sum_{\g=1}^{g-1}\Acal_P(R_\g)+\Acal_Q(R_g)+K^P\, | \,{\B})= $$
$$=-\frac{1}{2}\pm\log \Theta(K^P-\Acal(P_m))\pm\log[ \Si(P_m, Q_0)E^g(P_m, P)]-\frac{1}{4}\sum_{\a=1}^g\partial_{z_\a}\log \Theta(K^P-\Acal_P(Q))\pm v_\a(P_m)
$$
$$+\frac{1}{4\Theta(K^P-\Acal_P(Q))}\sum_{\a, \b=1}^g\partial^2_{z_\a z_\b}\Theta(K^P-\Acal_P(Q))v_\a(P_m)v_\b(P_m)
$$
$$
-\frac{1}{2}\sum_{\a=1}^g\partial_{z_\a}\log \Theta(K^P-\Acal_P(Q))\pm[\log E(Q, P_m)] v_\a(P_m)
$$
$$
=-\frac{1}{2}\pm\log \Theta(K^P-\Acal(P_m))\pm\log[ \Si(P_m, Q_0)E^g(P_m, P)]+\frac{\pm^2\Theta(K^P-\Acal(P_m))}{4\Theta(K^P-\Acal_P(P_m))}
$$
\begin{equation}\label{ochdlinn}
-\frac{1}{2}\sum_{\a=1}^g\partial_{z_\a}\log \Theta(K^P-\Acal_P(Q))\pm[\log E(Q, P_m)]v_\a(P_m)+o(1)
\end{equation}
as $Q\to P_m$.
The variation of remaining terms in the right hand side of (\ref{repC}) is much easier. One has
\begin{equation}\label{zero1}\lim_{R_1, \dots, R_g\to P}\partial_{\lambda_m} \sum_{\a<\b}\log E(R_\a, R_\b)=0\, ,\end{equation}
\begin{equation}\label{zero2}\lim_{R_1, \dots, R_g\to P}\partial_{\l_m}\sum_{\a=1}^g\log \Si (R_\a, P)=0
\end{equation}
\begin{equation}\label{formaQ}
\lim_{R_1, \dots, R_g\to P}\partial_{\l_m}\sum_{\a=1}^g\log E(Q, R_\a)=-\frac{g}{4}\left(
\pm\log\frac{E(Q, P_m)}{E(P, P_m)}
\right)^2
\end{equation}
due to (\ref{primeform}) and (\ref{sigmadiff}).
Now using (\ref{repC}), summing up (\ref{sigmadiff}), (\ref{ochdlinn} - \ref{formaQ}) and (\ref{denden}),
cleverly rearranging the terms (as Fay does on p. 59 of \cite{Fay92}) and sending $Q\to P_m$, we get
$$\partial_{\l_m}\cdiff(P)=\frac{1}{4}\frac{\pm^2\Theta(K^P-\Acal_P(P_m))}{\Theta(K^P-\Acal_P(P_m))}-\frac{1}{2}
\pm\log \Theta(K^P-\Acal_P(P_m))\pm\log [\Si(P_m, Q_0)E^g(P_m, P)]
$$
$$-\frac{1}{4}\pm^2\log E(P_m, P)+\frac{1}{2}\pm\log\Si(P_m, Q_0)\pm\log E(P_m, P)+
\frac{2g-1}{4}(\pm\log E(P_m, P))^2    
$$
$$
-\frac{1}{2}\left[\pm\log E(P_m, Q)\left(\sum_{\a=1}^g\partial_{z_\a}\log\Theta(K^P-\Acal_P(Q))v_\a(P_m)+
\pm\log[\Si(P_m, Q_0)E^g(P_m, P)]\right)
\right.$$
\begin{equation}\label{pochtifin}
\left. -\frac{1}{2}\frac{\pm^2 E(P_m, Q)}{E(P_m, Q)}-\sum_{\a, \b=1}^g\partial_{z_\a z_\b}^2\log\Theta(K^P-\Acal_P(P_m))v_\a(P_m)v_\b(P_m)\right]_{Q=P_m}\,.
\end{equation}
Due to (\ref{primas}), one has
$$\lim_{Q\to P_m}\pm\log E(P_m, Q)\left(\sum_{\a=1}^g\partial_{z_\a}\log\Theta(K^P-\Acal_P(Q))v_\a(P_m)+
\pm\log[\Si(P_m, Q_0)E^g(P_m, P)]\right)
$$
$$
=\lim_{x_m\to 0}\frac{1}{x_m}\left(\pm\log\frac{\Si(P_m, Q_0)E^g(P_m, P)}{\Theta(K^P-\Acal(P_m))}+
\sum_{\a, \b=1}^g\partial_{z_\a z_\b}^2\log\Theta(K^P-\Acal_P(P_m))v_\a(P_m)v_\b(P_m)x_m+O(x_m^2)\right)
$$
$$=\sum_{\a, \b=1}^g\partial_{z_\a z_\b}^2\log\Theta(K^P-\Acal_P(P_m))v_\a(P_m)v_\b(P_m)\;,$$
where we made use of the fact that the function
\begin{equation}\label{Rind}
R\mapsto \frac{\Si(R, Q_0)E^g(R, P)}{\Theta(K^P-\Acal_P(R))}
\end{equation}
for fixed $P$ is  holomorphic and single-valued  on $\L$ and, therefore,  a constant (thus the first term in the brackets vanishes).
Using (\ref{primas}) we see that
$$
\lim_{Q\to P_m}\frac{\pm^2 E(P_m, Q)}{E(P_m, Q)}=-\frac{1}{2}S_B(x_m)|_{x_m=0}\;.
$$
Thus, the last two lines of (\ref{pochtifin}) simplify to $-\frac{1}{8}S_B(x_m)|_{x_m=0}$.
Using the $R$-independence of expression (\ref{Rind}) once again, we may rewrite the first two lines of (\ref{pochtifin})
as
$$\frac{1}{4}\pm^2\log[\Si(P_m, Q_0)E(P_m, P)^{g-1}]-\frac{1}{4}(\pm\log[\Si(P_m, Q_0)E(P_m, P)^{g-1}])^2,$$
which coincides with $\frac{1}{8}S_{Fay}(x_m)|_{x_m=0}$ due to relation (\ref{proektivnaja}).
Formula (\ref{cmultdiff}) is proved. $\square$

\subsection{Dirichlet integral: variational formulas and holomorphic 
factorization}

 For the local parameter near the point at infinity
$\infty_n$ in this section we shall use the notation $\zeta_n:=1/\l$, which 
is the same as the parameter $x_{M+n}$ from the introduction.

\subsubsection{Definition of regularized Dirichlet integral}
\la{Diric}

Let us cut the branched covering $\L$ into $N$ sheets by a family of 
contours connecting ramification points $P_m$;
in addition, we dissect it along all $a$-cycles.
On each sheet of the covering $\L$ dissected in this way we can define 
a real-valued function
\begin{equation}
\label{potential}
\phi(P)=\log\left|\frac{\omega(P)}{d\pi(P)}\right|^2
\end{equation}
The difference of values of function $\phi$ on different sides of  
cycle $a_\a$ equals
$4\pi i(K_\a^P-\overline{K_\a^P})$.
The function $\phi$ is singular at all the points of the divisor $\Dcal$ (i.e. 
ramification points $P_1,\dots,P_M$ and points at
infinity $\infty_1,\dots,\infty_N$ of the branch covering $\L$) and at the point $P_0$. 
 The derivative $\p_\l\phi$ (where $\l=\pi(P)$) is holomorphic outside 
of the singularities of the 
function $\phi$ and does not change under tracing along the 
$a$-cycles.

\begin{lemma}
Projective connection (\ref{Sfa}) is related to function $\phi$ 
(\ref{potential}) everywhere outside of the 
 divisor $\Dcal$ as follows:
\begin{equation}\label{SvFay}
\Sfay(\l)=\phi_{\l\l}-\frac{1}{2}\phi_\l^2
\end{equation}
\end{lemma}
The proof of this lemma is a simple standard computation.

In terms of the function $\phi$ we define $M$ functions  
$\phi^{int}(x_m)$, which are analytic  in 
corresponding neighbourhoods of the ramification points $P_m$, as 
follows:
\begin{equation}\label{interplay}
e^{\phi^{int}(x_m)}|dx_m|^2=e^{\phi(P)}|d\l|^2\;;
\end{equation}
in analogy to (\ref{SvFay}) we get $\Sfay(x_m)=\phi^{int}_{x_m 
x_m}-\frac{1}{2}(\phi^{int}_{x_m})^2$.

Similarly, in a neighbourhood of any point at infinity $\infty_n$ we 
define   the function
 $\phi^{\infty}(\zeta_n)$
of the local parameter $\zeta_n$  by the equality 
$e^{\phi^{\infty}}|d\zeta_n|^2=
e^{\phi}|d\l|^2$. The projective connection $\Sfay$ in the parameter 
$\zeta_n$ coincides with 
$\phi^{\infty}_{\zeta_n\zeta_n}-\frac{1}{2}(\phi^{\infty}_{\zeta_n})^2$.

Using the interplay  between the functions $\phi$, $\phi^{int}$ and 
$\phi^{\infty}$,
 we find the following asymptotics
near the ramification points $P_m$ and the poles $\infty_n$:
\begin{equation}\label{aspoint}
|\phi_\l(P)|^2=\frac{1}{4}|\l-\l_m|^{-2}+O(|\l-\l_m|^{-3/2}) \ \ \ 
\text{as} \ P\to P_m
\end{equation}
and
\begin{equation}\label{aspole}
|\phi_\l(P)|^2=4|\l|^{-2}+O(|\l|^{-3}) \ \ \ \text{as} \ \  
P\to\infty_n\;.
\end{equation}

At the zero $P_0$ of the differential $\omega$ one gets 
\begin{equation}\label{aszero}
|\phi_\l(P)|^2=4(g-1)^2|\l-\l_0|^{-2}+O(|\l-\l_0|^{-1})  \ \ \text{as} 
\  \  P\to P_0.
\end{equation}
where $\l_0:=\pi(P_0)$.

These asymptotics enable us to introduce the following regularized 
Dirichlet integral
\begin{equation}\label{Dir}
{\mathbb 
D}=\frac{1}{\pi}\text{reg}\int_{\L}|\phi_\l|^2\hatdl=\frac{1}{\pi}\lim_{\rho\to 0}
\left\{I_\rho+\pi(M+8N+8(g-1)^2)\log\rho\right\},
\end{equation}
where $\hatdl =|d\l\wedge d\bar\l|/2$ and
\be
I_\rho=\sum_{n=1}^N\int_{\L^{(n)}_\rho}|\phi_\l|^2\hatdl\;.
\la{Irho}
\ee
Here $\L^{(n)}_\rho$ is the sub-domain of the $n$-th sheet of covering 
$\L$ obtained by cutting off the (small) discs of radius $\rho$
centred at the ramification points and (if applicable) $P_0$ from the 
(large) disc $\{\l\in \L^{(n)}: |\l|<1/\rho\}$.

\subsubsection{Holomorphic factorization of Dirichlet integral}

The following theorem shows how to compute the Dirichlet integral 
(\ref{Dir}) in terms of the local data at the
points of divisor $\Dcal$ and points $P_0$ and $Q_0$.

\begin{theorem}\label{derDir} 
The regularized Dirichlet integral admits the following 
representation:
\begin{equation}\label{Dirhol}
{\mathbb D}=\log\left|\frac{\sig^{4-4g}(P_0, 
Q_0)\prod_{m=1}^M\omega(P_m)}{\prod_{n=1}^N\omega^2(\infty_n)}
\exp\left\{4\pi i \langle \rb, K^{P_0}\rangle \right\}
\right|^2-2M\log2,
\end{equation}
where vector $\rb$ has integer components given by 
\be
2\pi 
r_\a:=\text{Var}|_{a_\a}\,\left\{\text{Arg}\,\f{\omega(P)}{d\pi(P)}\right\}\;.
\la{qa}\ee
\end{theorem}

{\bf Proof.}
Applying the Stokes theorem, we get
\begin{equation}\label{Qrho}
I_\rho=\frac{1}{2i}\left\{\sum_{m=1}^M\oint_{P_m}+
\sum_{n=1}^N\oint_{\infty_n}+\oint_{P_0}+\sum_{\a=1}^g\int_{a_\a^+\cup 
a_\a^-}\right\}
\phi_\l\phi\,d\l,
\end{equation}
  
Here $\oint_{P_m}$ and $\oint_{P_0}$ are integrals over clock-wise 
oriented circles of 
radius $\rho$ around the points $P_m$ and $P_0$
(it should be noted that each of the points $P_m$ belongs to two sheets 
simultaneously and, therefore, the  integration
in $\int_{P_m}$ goes over two circles). The $\oint_{\infty_n}$ denotes 
the integral over the counter-clock-wise oriented circle
of radius $1/\rho$  on the $n$-th sheet; $a_\a^+$ and $a_\a^-$ are 
different shores of the cycle $a_\a$ with the 
 opposite orientation. 
One has the equality
\begin{equation}\label{cycles}
\frac{1}{2i}\int_{a_\a^+\cup a_\a^-}\phi_\l\phi\,d\l=\pi r_\a\log|\exp 
4\pi i K^{P_0}_\a|^2\;.
\end{equation}
We note that $\phi_\l=\partial_\l\log(\omega(P)/d\pi(P))$, where the 
function $\omega(P)/d\pi(P)$ is single-valued on the  
cycle $a_\a$;  since
the $a$-cycles are assumed not to contain the point $P_0$, function  
$\omega(P)/d\pi(P)$ does not vanish on $a_\a$.

We have also

\begin{equation*}
\frac{1}{2i}\oint_{P_m}=\frac{1}{2i}\oint_{|x_m|=\sqrt{\rho}}\left\{
\phi^{int}_{x_m}\frac{1}{2x_m}-\frac{1}{2x_m^2}\right\}\{\phi^{int}-2\log|x_m|-2\log2\}2x_m\,dx_m=
\end{equation*}
\begin{equation}\label{Pm}
=\pi\phi^{int}(x_m)|_{x_m=0}-2\pi\log2-\pi\log\rho+o(1),
\end{equation}

\begin{equation*}
\frac{1}{2i}\oint_{\infty_n}=\frac{1}{2i}\oint_{|\l|=1/\rho}\left\{
-\phi_{\zeta_n}^\infty\l^{-2}-\frac{2}{\l}\right\}\{\phi^{\infty}-4\log|\l|\}\,d\l=
\end{equation*}
\begin{equation}\label{inftyn}
=-2\pi\phi^{\infty}(\zeta_n)|_{\zeta_n=0}-
8\pi\log\rho+o(1),
\end{equation}

and

\begin{equation*}
\frac{1}{2i}\oint_{P_0}=\frac{1}{2i}\oint_{|\l-\l_0|=\rho}\log|\sigma^2(P_0, 
Q_0)(\l-\l_0)^{2g-2}\{1+O(\l-\l_0)\}|^2
\times
\end{equation*}
\begin{equation}\label{point}
\times\left(
\frac{2g-2}{\l-\l_0}+O(1)\right)d\l=
-\pi\log|\sigma^{4g-4}(P_0, Q_0)|^2-8\pi(g-1)^2\log\rho+o(1),
\end{equation}
as $\rho\to 0$.
These asymptotics together with (\ref{Dir}) and (\ref{cycles}) imply 
(\ref{Dirhol}).
$\square$

\subsubsection{Variational formulas for Dirichlet integral}

From now on we assume that the projections $\pi(P_0)$ and $\pi(Q_0)$  
of the points $P_0$ and $Q_0$ from (\ref{Fdiff}) 
are independent of $\{\l_m\}$.

\begin{theorem}\label{DirFay}
The variation of the regularized Dirichlet integral ${\mathbb D}$ 
(\ref{Dir}), (\ref{Dirhol}) with respect to branch points $\l_m$
is given by values of projective connection (\ref{Sfa}) at the 
ramification points $P_m$ i.e. 
\begin{equation}
\frac{\partial {\mathbb D}}{\partial \l_m}=\Sfay(x_m(P))|_{P=P_m}, \ \ 
\ m=1, \dots, M\;.
\end{equation} 
\end{theorem}

We start from the following 

\begin{lemma}\label{le}
On every sheet of the covering $\L$, 
 dissected in addition along all $a$ and $b$-cycles, the   derivatives 
of  function 
$\phi(P)$ with respect to $\l$ and $\l_m$ are related as follows:
\begin{equation}\label{Ahl}
\phi_{\l_m}+F_m\phi_\l+(F_m)_\l=0\;, \ \ \ m=1, \dots, M,
\end{equation}
where functions $F_m(P)$ are defined on the dissected covering $\L$ as 
follows: 
$$F_m(P)=-\frac{\Ucal(P)_{\l_m}}{\Ucal(P)_\l}\;,$$
and $\Ucal(P)=\int_{P_0}^P\omega$.
Near ramification points and points at infinity the functions $F_m$ 
have the following asymptotics:
$$F_m(P)=O(|\l|^2)\;,\hskip0.6cm {\rm as}\hskip0.6cm P\to \infty_n\;, 
$$ 
$$F_m(P)=\delta_{lm}+o(1)\;,\hskip0.6cm {\rm as}\hskip0.6cm P\to 
P_l\;,$$ 
where $\delta_{lm}$ is the Kronecker symbol. 
\end{lemma}

The proof of relation
(\ref{Ahl}) can be obtained by direct differentiation (one needs to use 
the fact that the map $\Ucal$ depends on 
 $\{\l_m\}$ holomorphically). The proof of the 
asymptotical behaviour of the functions $F_m$  essentially repeats 
Lemma 5 from \cite{KokKor1}.

Considering the exact differentials $d((F_m)_\l\phi)$, 
$d(F_m\phi_\l\phi)$,
$d(F_m\phi_\l)$ and making use of (\ref{Ahl}), we get the following 

\begin{corollary}
The following two 1-forms are exact:
\begin{equation}\label{e1}
\{(\phi_\l\phi)_{\l_m}d\l \}-\{ 
F_m|\phi_\l|^2d\bar\l+(F_m)_\l\phi_{\bar\l}d\bar\l\;\}\;,
\end{equation}
and
\begin{equation}\label{e2}
\{F_m|\phi_\l|^2 d\bar\l-(F_m)_\l\phi_\l 
d\l+(F_m)_\l\phi_{\bar\l}d\bar\l\}-
\{F_m(2\phi_{\l\l}-\phi_\l^2)d\l+\phi\phi_{\l\l_m}d\l\}\;.
\end{equation}
\end{corollary}

{\it Proof of theorem \ref{DirFay}} (the idea of this proof, 
including lemma \ref{le}  goes back  to  \cite{Prokoly}).

By (\ref{Qrho}) we get
\begin{equation*}
\frac{\partial I_\rho}{\partial \l_m}=\frac{1}{2i}
\left\{\oint_{P_m}\left[(\phi_\l\phi)_{\l_m}+(\phi_\l\phi)_\l\,\right]d\l+
\left(\sum_{l\neq 
m}\oint_{P_l}+\sum_{n=1}^N\oint_{\infty_n}+\oint_{P_0}\right)
(\phi_\l\phi)_{\l_m}\,d\l
\right\}+
\end{equation*}
\begin{equation}\label{proof1}
+\frac{1}{2i}\sum_{\a=1}^g\int_{a_\a^+\cup 
a_\a^-}(\phi_\l\phi)_{\l_m}\,d\l\;.
\end{equation}
(One may assume that the projections of basic cycles $\pi(a_\a)$ are 
independent of $\{\l_m\}$.)

Using the holomorphy of $(F_m)_\l\phi_\l$ and the relation 
$(\phi_\l\phi)_\l\,d\l=d(\phi_\l\phi)-\phi_\l\phi_{\bar\l}d\bar\l$,
we rewrite the r. h. s. of (\ref{proof1}) as
$$\frac{1}{2i}\Big[-\oint_{P_m}|\phi_\l|^2d\bar\l+\left\{
\sum_{l=1}^M\oint_{P_l}+\sum_{n=1}^N\oint_{\infty_n}+\oint_{P_0}\right\}
\{F_m|\phi_\l|^2-(F_m)_\l\phi_\l\,d\l+(F_m)_\l\phi_{\bar\l}\,d\bar\l\}-$$
\begin{equation}\label{proof2}
-\sum_{\a=1}^g\left[\int_{a_\a^+\cup a_\a^-}+\int_{b_\a^+\cup 
b_\a^-}\right](F_m)_\l\phi_\l\,d\l+\sum_{\a=1}^g\int_{a_\a^+\cup 
a_\a^-}(\phi_\l\phi)_{\l_m}\,d\l\Big].
\end{equation}

By means of the asymptotical expansions of the integrands at the 
ramification points and points $\infty_n$
(the asymptotics from Lemma \ref{le} play here a central role;
cf. the proof of Theorem 4 in 
\cite{KokKor1}) one gets the relation
\begin{equation*}
\frac{1}{2i}\left\{\sum_{l=1}^M\oint_{P_l}+\sum_{n=1}^N\oint_{\infty_n}\right\}\{
F_m|\phi_\l|^2d\bar\l
-(F_m)_\l\phi_\l d\l+(F_m)_\l\phi_{\bar\l}\,d\bar\l
\}=
\end{equation*} 
\begin{equation}\label{proof3}
=\frac{1}{2i}\oint_{P_m}|\phi_\l|^2\,d\bar\l-\frac{3\pi}{4}\sum_{l=1}^M\frac{d^2F_m}{(dx_l)^2}\Big|_{x_l=0}+o(1)
\end{equation}
as $\rho\to 0$.

By (\ref{e2}) we have
\begin{equation}\label{proof4}
\frac{1}{2}\oint_{P_0}
F_m|\phi_\l|^2 d\bar\l
-(F_m)_\l\phi_\l d\l+(F_m)_\l\phi_{\bar\l}\,d\bar\l=
\frac{1}{2i}\oint_{P_0} F_m(2\phi_{\l\l}-(\phi_\l)^2)d\l+o(1).
\end{equation}

The Cauchy theorem, the asymptotical expansions at $P_l$ and $\infty_n$ 
and relation (\ref{SvFay}) imply that
\begin{equation*}
0=\frac{1}{2i}\left\{\sum_{l=1}^M\oint_{P_l}+\sum_{n=1}^N\oint_{\infty_n}+\oint_{P_0}+
\sum_{\a=1}^g\left[\int_{a_\a^+\cup a_\a^-}+\int_{b_\a^+\cup 
b_\a^-}\right]\right\}
F_m(2\phi_{\l\l}-(\phi_\l)^2)\,d\l=
\end{equation*}
\begin{equation}\label{proof5}
=-\frac{3\pi}{4}\sum_{l=1}^M\frac{d^2F_m}{(dx_l)^2}\Big|_{x_l=0}+\pi 
\Sfay (x_m)\Big|_{x_m=0}+\frac{1}{2i}
\left\{\oint_{P_0}+\sum_{\a=1}^g\left[\int_{a_\a^+\cup a_\a^-}
+\int_{b_\a^+\cup b_\a^-}\right]
\right\}
F_m(2\phi_{\l\l}-(\phi_\l)^2)\,d\l+o(1)
\end{equation}
(cf. \cite{KokKor1}, Lemma 6).
Observe that
 $$\int_{a_\a^+\cup a_\a^-}(\phi_\l\phi)_{\l_m}\,d\l=
\int_{a_\a^+\cup 
a_\a^-}[(2\phi_{\l\l}-(\phi_\l)^2)F_m+(F_m)_\l\phi_\l]\,d\l$$
due to (\ref{e2}) and an obvious equality
$$\int_{a_\a^+\cup a_\a^-}\phi\phi_{\l\l_m}\,d\l=0.$$
Similarly,
$$\int_{b_\a^+\cup 
b_\a^-}[(2\phi_{\l\l}-(\phi_\l)^2)F_m+(F_m)_\l\phi_\l]\,d\l=0,$$
due to the equality
$$\frac{\partial}{\partial\l_m}\int_{b_\a^+\cup 
b_\a^-}\phi_\l\phi\,d\l=0.$$
To finish the proof it remains to collect together equations 
(\ref{proof1})-(\ref{proof5}),
and make use of the fact that all $o(1)$ in the above equalities are 
uniform with respect to $(\l_1, \dots, \l_M)$ belonging to 
a compact neighbourhood of the initial point $(\l_1^0, \dots, \l_M^0)$. 
$\square$

\subsection{Calculation of the  tau-function}

\begin{theorem}\la{main1}
The isomonodromic tau-function of Frobenius manifold structure on the Hurwitz space
$H_{g,N}(1,\dots,1)$
is given by the following expression, which 
is independent of the choice of the points $P_0$ and $Q_0$:
\begin{equation}\label{prel1}
\tau^{-6}=\frac{\{\Si(P_0, Q_0)\}^{2-2g}e^{2\pi 
i\langle\rb,K^{P_0}\rangle}}{\cdiff^4(P_0)(d\pi(P_0))^{g-1}}
\prod_{k=1}^{M+N} \{\Si(D_k,Q_0)\}^{d_k} \{E(D_k,P_0)\}^{(g-1)d_k}\;,
\end{equation}
where the integer vector $\rb$ is defined as follows:
\be
 \Acal(\Dcal)  +2K^{P_0}+\B\rb+\sb=0\;;
\la{defind1}
\ee 
the initial point of the Abel map coincides with $P_0$ and all the 
paths are chosen inside the same fundamental polygon $\Lhat$.
\end{theorem}

{\bf Proof.} Expression (\ref{prel1}) is an immediate corollary of 
theorems \ref{derDir}, \ref{DirFay} and formula (\ref{cmultdiff}).
The only thing
one needs to check is the coincidence of the vector $\rb$ defined by 
formula (\ref{qa}) with the vector $\rb$ defined by (\ref{defind1}).
This coincidence is easy to prove for $g\geq 2$. Namely, we know that 
(\ref{prel1}), where the components of the vector $\rb$ are
given by (\ref{qa}), gives the  tau-function (\ref{deftau}). 
Therefore, as $P_0$ encircles the basic $b$-cycle,
the tau-function can only gain a $\{\l_m\}$-independent factor. 
Computing the monodromy of expression (\ref{prel1})
along cycle $b_\a$, we see that this implies (\ref{defind1}) unless 
$g-1\neq 0$. For $g=1$ the relation (\ref{prel1})
follows from the formula for the tau-function  which was found 
in \cite{KokKor1}.

Now, since (\ref{defind1}) is proved, we can show that expression 
(\ref{prel1}) is independent of $P_0$ and $Q_0$.
Simple counting of tensor weight shows that expression (\ref{prel1}) is 
a $0$-differential  with respect to each argument
 $P_0$ and $Q_0$, which is, moreover, free of singularities. Due to 
relation (\ref{defind1}) and multiplicative properties 
of the differentials $\cdiff$ and $\Si$ we can also check that it is 
single-valued on $\L$ with respect to each of these 
arguments, and, therefore, is independent of both of them.

$\Box$

To transform expression (\ref{prel1}) further to the form 
(\ref{tauint}) we shall use the following two lemmas

\begin{lemma}
The fundamental domain $\Lhat$ of the Riemann surface $\L$  can always be chosen such that
$\Acal(\Dcal)+2K^P=0$.
\end{lemma}
%This lemma is valid since all zeros of the differential $d\pi$ are simple; actually, it is sufficient to
%have only one simple zero of $d\pi$ to make it true.

{\it Proof.} 
For an arbitrary choice of the fundamental domain  the vector $\Acal(\Dcal)+2K^P$
coincides with $0$ on the jacobian of the surface $\L$ i.e. there exist two integer vectors ${\bf r}$ and
${\bf s}$ such that 
$$
\Acal(\Dcal)+2K^P+{\bf B}{\bf r}+ {\bf s}=0
$$
Consider the point $P_1\in \Dcal$; according to our assumptions this is a simple zero of $d\pi$.
By a smooth deformation of a cycle $a_\a$ within a given homological class we can stretch it
 in such a way that the point $P_1$ crosses this cycle; two possible
directions of this crossing correspond to the jump of the component
${\bf r}_\a$ of the vector ${\bf r}$ to $+1$ or $-1$. Similarly, if we
deform a cycle $b_\a$ in such a way that it gets crossed by the point
$P_1$, the component ${\bf s}_\a$ of the vector ${\bf s}$ also jumps
with $+1$
or $-1$ depending on the direction of the crossing. Repeating such procedure, we get the fundamental domain where 
${\bf r}={\bf s}=0$. $\Box$

From the proof it is clear that even a stronger statement is true: the choice of the fundamental domain
such that $\Acal(\Dcal)+2K^P=0$ is possible if at least one of the ramification points is simple.

\begin{lemma}
Assume that the fundamental domain $\Lhat$ is chosen in such a way that
\be
\Acal(\Dcal)+2K^P=0.
\la{defind2}
\ee 
Then for any two points $P,Q\in\L$ the function $\Si(P,Q)$ can be written as 
follows in terms of prime-forms:
\be
\Si^2(P,Q)=\f{d\pi(P)}{d\pi(Q)}\prod_{k=1}^{M+N}\left(\f{E(D_k,Q)}{E(D_k,P)}\right)^{d_k}
\la{siPQ}
\ee
\end{lemma}
{\it Proof.} Consider the expression 
\be
{(d\pi(P))^{1-g}}\cdiff^{-2}(P)\prod_{k=1}^{M+N} E^{d_k(g-1)}(D_k,P)\;. 
\la{expre}
\ee
Summing up the tensor weights of all ingredients of this  expression, 
we see that this  is a 
 $0$-differential with respect to $P$. Moreover, it is single-valued 
under tracing along all
the basic cycles (this can be easily checked using multiplicative 
properties of the prime-forms and $\cdiff(P)$), and  
does not have either zeros or poles on $\L$ (the poles and zeros 
induced by the prime-forms  are cancelled  by the poles 
and zeros of $d\pi(P)$). Therefore, expression (\ref{expre}) is 
independent of the point $P$.
Taking its ratio at arbitrary two points $P$ and $Q$ and using  
expression  (\ref{Csi}) of $\Si(P,Q)$ 
in terms of ratio of the differential $\cdiff(P)$ at these two points, 
we get (\ref{siPQ}).
$\Box$

Now, choosing in  formula (\ref{prel1}) $P_0=Q_0$ and expressing 
$\Si(D_k,P_0)$ in terms of the prime-forms using
(\ref{siPQ}), we get expression (\ref{tauint}) for the isomonodromic
tau-function of Hurwitz Frobenius manifolds 
stated in the introduction.

\begin{remark}\rm
It is natural to expect that once the final expression (\ref{tauint})
is known, the validity of defining equations (\ref{deftau}) can be proved
by a straightforward computation without using the technique of
variation and holomorphic factorization of the Dirichlet
integral. Such straightforward proof is indeed possible in the genus
zero and genus one cases
\cite{KokStr}; however, surprisingly enough, it seems to be more
technical than the indirect proof using the technique of Dirichlet
integral.
Therefore, although we believe that the direct proof  exists also in
the higher genus case,
probably, it does not lead to a significant simplification of the
proof given here. 
\end{remark}

\begin{remark}\rm
For the stratum $H_{2,g}(1,\dots,1)$, which consists of hyperelliptic
Riemann surfaces
$\nu^2=\prod_{j=1}^{2g+2}(\l-\l_j)$
 with $M=2g+2$ simple branch points, the tau-function
$\tau$ was computed in \cite{KitKor} in the following form:
\be
\tau={\rm det}{\bf A} \prod_{j<k,\;j,k=1}^{2g+2} (\l_j-\l_k)^{1/4}
\la{tauhyp}
\ee
where ${\bf A}_{\a\b}=\oint_{a_\a}\f{\l^{\b-1}}{\nu}$ is the matrix of
$a$-periods of non-normalized holomorphic abelian differentials on
$\L$.
To verify that expression (\ref{tauint}) in hyperelliptic case gives
rise to (\ref{tauhyp}) we need to use the representation (\ref{repC})
for the differential ${\cal C}(P)$, together with the formula
(\ref{siPQ}) for the differential ${\bf s}(P,Q)$ and assume that $g+1$ arbitrary
points $Q,R_1,\dots,R_g$ tend to $g+1$ different branch points (say, $\l_1,\dots,\l_{g+1}$).
The theta-function entering (\ref{repC}) can be then computed via
Thomae formula in terms of ${\rm det}{\bf A}$ and pairwise differences
of the branch points. The ${\rm det} ||v_\a(R_\b)||$ is also easily
represented in the same terms. Collecting all appearing contributions,
we arrive at (\ref{tauhyp}).
\end{remark}

\subsubsection{Genus 1 case}

The differential $\cdiff(P)$ in elliptic case does not depend on $P$ 
(\cite{Fay92}, p.21): 
$$\cdiff=\eta^3(\B)e^{-\pi i\B/4}\;,$$
where ${\eta}({\B})$ is the Dedekind eta-function.
The differential $\Si(P,Q)$ is given by
\be
\Si(P,Q)=\exp\left\{\pi i\int_Q^P 
v\right\}\frac{\sqrt{v(P)}}{\sqrt{v(Q)}}
\ee
Substituting these expressions to (\ref{prel1}) and taking into account 
(\ref{defind1}), we get the following expression:
\be
\tau=\eta^2(\B)\prod_{k=1}^{M+N} \{v(D_k)\}^{-{d_k}/12}\;,
\ee
where according to our usual conventions  
$v(D_k):=v(P)/dx_k(P)|_{P=D_k}$, $k=1,\dots,M+N$.
This formula was independently proved in \cite{KokKor1}; this confirms 
correctness of the choice of integer $\rb$  
in (\ref{prel1}) for $g=1$.

\subsection{Tau-function for an arbitrary stratum of Hurwitz space}

Here we briefly 
consider the general case, when the critical points and poles of 
function $\pi(P)$ have arbitrary multiplicities.
Such tau-function arises for general Hurwitz Frobenius manifolds from (\cite{D})
and in the problem of computation of the subleading term in the large $N$ expansion of
the partition function in hermitian two-matrix model \cite{EKK}
(in this case the multiplicities of poles of $\pi(P)$ can be 
arbitrary, and the branch points are simple).
In the problem of computation of 
isomonodromic tau-function 
corresponding to Riemann-Hilbert problem with arbitrary permutation 
monodromies \cite{Kor03} the 
multiplicities of the critical points can be arbitrary. 
As before, denote the branch points of the branched covering $\L$  by 
$P_1,\dots,P_M$ and assume that they
have multiplicities $d_1,\dots,d_M$; the orders of the poles 
$\infty_1,\dots,\infty_L$ of $\pi$ we denote by 
$d_{M+1}-1,\dots,d_{M+L}-1$, respectively. One has 
$N=\sum_{s=1}^{L}d_{M+s}$.
Then divisor $\Dcal:=(d\pi)$ can be formally written in the same  form 
as before:
\be
\Dcal=\sum_{k=1}^{M+L} d_k D_k
\la{Dmult}
\ee
where $D_m:=P_m$, $m=1,\dots,M$ and $D_{M+s}=\infty_s$, $s=1,\dots, 
L$.
The genus of the Riemann surface $\L$ is given in this case by the 
formula:
\be
g=\f{1}{2}\sum_{m=1}^M d_m - \sum_{s=1}^L d_{M+s} +1
\ee
The definition (\ref{deftau}) generalises as follows: 
\be
\f{\p}{\p \l_m}\log\tau=-\f{1}{6(d_m-1)!(d_m+1)} 
\left(\f{d}{dx_m(P)}\right)^{d_m-1}S_B(x_m(P))\Big|_{P_m=P}\;,
\hskip0.8cm m=1,\dots,M\;.
\la{daftau1}
\ee
Compatibility of the system (\ref{daftau1}) follows from Schlesinger 
equations \cite{Kor03}; it was checked directly 
in \cite{KokKor1}
using Rauch variational formulas.
By using the same technique as in the case of simple branch points and 
infinities one can verify that the 
formula (\ref{tauint}) stated in the introduction remains valid in the 
general case after substitution of
corresponding multiplicities $d_k$ and genus $g$. We don't present the 
proof here since it does not contain any
new essential ideas in comparison with the case of simple poles and 
zeros.

\section{Applications of the tau-function of Hurwitz Frobenius manifolds}

\subsection{G-function of Frobenius manifolds}

Fix a stratum $H_{g, N}(k_1, \dots, k_L)$ of Hurwitz space, for which 
all the critical points of function $\pi(P)$ are simple, but 
infinities  have  arbitrary multiplicities $k_1,\dots,k_L$ (for simple 
infinities all $k_s=1$). 
Then the divisor $\Dcal$ (\ref{Dmult}) of the 
differential $d\pi$ looks as follows:
\be\la{DF}
\Dcal_{Frob}=\sum_{m=1}^M P_m - \sum_{s=1}^L (k_s+1)\infty_s
\ee
i.e. $d_m=1$ for $m=1,\dots,M$ and $d_{M+n}=k_n+1$ for $n=1,\dots, L$.
The structures of Frobenius manifold on any 
 Hurwitz space of this type  were introduced by Dubrovin \cite{D}.
 We refer to \cite{D} or \cite{Manin} and recent papers \cite{Vaska1,Vaska2}
 for definition of all ingredients 
(Frobenius algebra, prepotential, canonical and flat coordinates,
Darboux-Egoroff metrics, $G$-function) of this construction. 
Here we shall only discuss the $G$-function (genus one free energy of 
Dijkgraaf and Witten), 
which gives a solution of Getzler equation
(for classes of Frobenius manifolds related to quantum cohomologies the 
$G$-function is the generating function
of elliptic Gromov-Witten invariants).

Recall that each Frobenius structure on 
the Hurwitz space corresponds to a so-called primary differential 
$\phi$ on the covering $\L$.
The arising Frobenius manifold will be denoted by $M_\phi$.
In \cite{DZ2, DZ3} it was found the following expression for 
the $G$-function of an  arbitrary semisimple Frobenius manifold:
\begin{equation}\label{G-f}
G=\log\left(\frac{\tau_I}{J^{\frac{1}{24}}}\right).
\end{equation}
where $\tau_I$ is the Dubrovin's  tau-function associated to an 
arbitrary semisimple Frobenius manifold (in \cite{D} $\tau_I$ is called the
``isomonodromic tau-function'', although, as was shown in \cite{KokKor2},
  it is related to the original definition of
Jimbo-Miwa, which we follow here, by $\tau_I=\tau^{-1/2}$); $J$ is the
Jacobian  of the transformation
 from the flat to the canonical coordinates.

Using the known expression of Jacobian $J$ in terms of diagonal 
coefficients of Darboux-Egoroff metric
(see, e. g., \cite{DZ2}), we get the following
\begin{theorem}
The $G$-function of the Frobenius manifold $M_\phi$ can be expressed as 
follows
\begin{equation}\label{GGTT}
G=-\frac{1}{2}\log\tau-\frac{1}{48}\sum_{m=1}^M\log\,{\rm 
Res}\,_{P_m}\frac{\phi^2}{d\l},
\end{equation}
where $\tau$ is the tau-function on the Hurwitz space $H_{g, 
N}(k_1, \dots, k_L)$ given by (\ref{tauint}), with
the divisor $\Dcal$ given by (\ref{DF}).
\end{theorem}

\subsection{Genus one free energy of hermitian two-matrix model}

Another application of the tau-function (\ref{tauint}) is in the
theory of hermitian one- and two-matrix models \cite{EKK}. Consider the partition
function of hermitian two-matrix model
\be
e^{-N^2F}:= \int dM_1 dM_2 e^{-N\tr\{V_1(M_1)+V_2(M_2)-M_1 M_2\}}\;.
\la{part}
\ee
where  the integration goes over all independent matrix entries of
$N\times N$ hermitian matrices $M_1$ and $M_2$; $V_1$ and $V_2$ are
two polynomial potentials (sometimes it is convenient to consider $V_1$ 
and $V_2$ as infinite power series).
The expansion $F=\sum_{G=0}^{\infty} N^{-2G} F^{G}$ as  $N\to\infty$
(so-called ``genus expansion'') plays
an important role in the theory, since the coefficients  $F^{G}$
appear both
in statistical physics (Ising model) as well as in  enumeration of genus $G$ graphs (see for
example \cite{DiF}). 
If polynomials $V_1$ and $V_2$ are of even degree with positive
leading coefficients, then asymptotically, as $N\to \infty$, the main
contribution to the partition function (\ref{part}) is given by the
matrices whose eigenvalues are concentrated in a finite set of
intervals. The intervals filled by the eigenvalues of the matrix $M_1$ lie
around the minima of the potential $V_1$; the eigenvalues of the
matrix $M_2$ fill the intervals around the minima of the potential $V_2$.

The intervals supporting eigenvalues of matrices $M_1$ and $M_2$
correspond to the so-called spectral algebraic curve $\L$, defined by
equation
\be
(V_1'(x)-y)(V_2'(y)-x) -\Pcal^{0}(x,y)+1=0
\la{spcurve}
\ee
 where the polynomial of two variables 
$\Pcal^{0}(x,y)$ is the zeroth order term in $1/N^2$ expansion of the polynomial  
\be
\Pcal(x,y):=\f{1}{N}\Big\langle \tr\f{V_1'(x)-V_1'(M_1)}{x-M_1}\f{V_2'(y)-V_2'(M_2)}{y-M_2}\Big\rangle\;;
\la{Pxy}
\ee
(the notation $\langle Q (M_1,M_2) \rangle$ is used to define the
expectation value of any functional $Q$ of the matrices  $M_1$ and
$M_2$ with respect to the integration measure in (\ref{part})).
The branch cuts of the spectral curve $\L$ corresponding to projection
of $\L$ on the $x$-plane coincide with the intervals supporting
eigenvalues of $M_1$ in the limit $N\to \infty$; the branch cuts
corresponding to projection of $\L$ on $y$-plane are the intervals
supporting the eigenvalues of $M_2$.

The equations for derivatives of the functions
$F^{G}$ with respect to coefficients of polynomials $V_{1,2}$ arise as a
corollary of a reparametrization invariance of the matrix integral
(\ref{part}) (the so-called ``loop equations'').
In particular, the formula for the leading order term $F^0$ (``genus
zero free energy'') in terms of standard holomorphic objects
associated to the spectral curve $\L$ was
derived in \cite{Bertola}. 
Results of \cite{Bertola}, together with \cite{ChekMir}, show, that $F_0$
satisfies so-called generalised WDVV equations, together with a quasi-homogeneity
equation, thus indicating the existence of a close link between the large
$N$ limit of hermitian matrix models and the theory of Frobenius
manifolds. Further confirmation of this link was obtained in
\cite{EKK} where it was shown that the genus one contribution $F^1$ to the
free energy is given by the formula
\be
F^{1}= \f{1}{2}  \log\tau   +    \f{1}{48}\log\left\{
(v_{d_2+1})^{1-\f{1}{d_2}}\prod_{m=1}^{M} \res|_{P_m}\f{(d y)^2}{d x} \right\}+ C
\la{F1int}
\ee
where $v_{d_2+1}$ is the highest order coefficient of 
the polynomial $V_2$; $P_1,\dots,P_M$ are  zeros of the differential $dx$ on
the spectral curve (i.e. the branch points of the spectral curve
realized as a covering of the $x$-plane), which are
assumed to be simple; $\tau$ is the isomonodromic tau-function
 of a Hurwitz Frobenius manifold associated to the
spectral curve (\ref{spcurve}). The formula (\ref{tauint}) proved in
this paper gives an explicit expression for the genus one free energy (\ref{F1int}).
The one-matrix model appears when the degree of polynomial $V_2$
equals $2$; in this case the spectral curve (\ref{spcurve}) is
hyperelliptic.

A surprising similarity of the expression for the $G$-function (\ref{GGTT}) of Hurwitz
Frobenius manifolds with the expression for the genus one free energy
(\ref{F1int}) of Hermitian two-matrix models is an additional evidence
of existence of a close link between hermitian matrix models and 2d
topological field theories.

\subsection{Determinant of  Laplacian in Poincar\'e metric}

Here we consider an application of the tau-function (\ref{tauint}) to 
computation of the determinants of Laplacians 
on Riemann surface in the Poincar\'e metric (in the trivial line 
bundle); such determinants
 are defined in terms of the corresponding $\zeta$-functions as 
follows: ${\rm det}\Delta:=\exp\{-\zeta_\Delta'(0)\}$.
For elliptic case and the flat metric $|v|^2$ (where $v$ is holomorphic 
normalized differential) this determinant is given 
by the Ray-Singer formula (see \cite{Ray} and formula (\ref{detlap1}) below). Such explicit 
formula is absent for $g>1$ for Poincar\'e
metric, although variational formulas for ${\rm det}\Delta$ with 
respect to the moduli of the Riemann surface are well-known
(see, e. g., \cite{ZT}, or  \cite{Fay92} (formulae (5.4) and (4.58))).
        As it was shown in \cite{KokKor1},  these formulas imply the
following expression for the derivative of ${\rm det}\Delta$ with
respect to a simple branch point of the covering $\L$:
\be
\f{\p}{\p\l_m}\left\{\f{{\rm det}\Delta}{{\rm 
det}\Im\B}\right\}=-\f{1}{12}(S_B-S_{Fuchs})(P_m)\;,
\ee
where $S_B$ is the Bergman projective connection, and $S_{Fuchs}(P):=\{z(P),x(P)\}$ is 
the Fuchsian projective connection on $\L$, where $z(P)$ if the
fuchsian uniformization coordinate; $x(P)$ is a local parameter.

%Namely, the variation of the determinant under the variation of 
%conformal structure given by a Beltrami differential $\mu$
%looks as follows:
%\be
%\delta_{\mu}\left\{\f{{\rm det}\Delta}{{\rm 
%det}\Im\B}\right\}=-\f{1}{12\pi i}\int_{\L}(S_B-S_{Fuchs})\mu
%\ee
%where $S_B$ is the Bergman projective connection, and $S_{Fuchs}$ is 
%the Fuchsian projective connection on $\L$.

Using this variational formula, in \cite{KokKor1} it was obtained a 
formula which expresses ${\rm det}\Delta$ in terms
of the  tau-function of Hurwitz Frobenius manifolds. 
To formulate the  theorem which combines this result with formula 
(\ref{tauint}) we need to introduce a few new objects.

Let all the critical points and poles of the function $\pi$ be simple 
(this is sufficient for computation of ${\rm det}\Delta$
since on any Riemann surface we can find a meromorphic function with 
these properties).
For  $g>1$ the Riemann surface $\L$ is  biholomorphically equivalent to 
the quotient 
space ${\mathbb H}/\Gamma$, where
${\mathbb H}=\{z\in {\mathbb C}\,:\, \Im z>0\}$; $\Gamma$ is a strictly 
hyperbolic  Fuchsian group.
Denote by $\pi_F:{\mathbb H}\rightarrow \L$ the natural projection.
Let $x$ be a local parameter on $\L$. Introduce the standard  metric  
of the constant
curvature $-1$ on $\L$:
\begin{equation}\label{locmetr}
e^{\chi}|dx|^2=\frac{|dz|^2}{|\Im z|^2}\;,
\end{equation}
where $z\in {\mathbb H}$, $\pi_F(z)=P$, $x=x(P)$.

In complete analogy to constructions of Sec. \ref{Diric},
introduce the real-valued functions $\chi(\l)$, $\chi^{int}(x_m)$, 
$m=1, \dots, M$ and
$\chi_n^{\infty}(\zeta_n)$, $n=1, \dots, N$ by specifying  the local 
parameter  $x=\l$, $x=x_m$
and $x=\zeta_n$ (in a neighbourhood of the point at infinity of the 
$n$-th sheet) in (\ref{locmetr}) respectively.

Consider domains
$\L_\rho^{(n)}$ of $\L$ as in the integrals (\ref{Irho}).
(Recall that the domain
$\L_\rho^{(n)}$ is obtained from the $n$-th sheet of $\L$ by deleting 
small discs
around ramification  points belonging to this sheet, and the disc 
around infinity.)

Define the regularized Dirichlet integral analogous to (\ref{Dir}):
\begin{equation}\label{reg2}
{\mathbb D}_F:=\f{1}{\pi}\lim_{\rho\to 0}
\left( \sum_{n=1}^N\int_{\L_\rho^{(n)}}|\partial_\l\chi|^2\widehat{d\l}       
+(8N+M)\pi\log \rho\right).
\end{equation}
Define the function ${\mathbb S}_F$ by
\begin{equation}\label{Main2}
{\mathbb S}_F(\l_1, \dots, \l_M)=-\f{1}{12}{\mathbb D}_F-
\frac{1}{6}\sum_{m=1}^M\chi^{int}(x_m)\Big|_{x_m=0}
+\frac{1}{3}\sum_{n=1}^N\chi_n^\infty(\zeta_n)\Big|_{\zeta_n=0}\; ;
\end{equation}

Now we are in a position to formulate the following
\begin{theorem}\label{det}
Consider the Hurwitz space $H_{g,N}(1,\dots,1)$. Let the pair 
$(\L,\pi)$ belong to $H_{g,N}(1,\dots,1)$.
Then the determinant of the Laplace operator on $\L$ (acting in the 
trivial line bundle)
in Poincar\'e metric  is given by the following expression:
\begin{equation}
{\rm det}\,\Delta=c_{g,N}\,\{{\rm det}\,\Im\B\} \, e^{{\mathbb S}_F} \,  
|\tau|^2.
\la{detlap}
\end{equation}
where $c_{g,N}$ is a constant independent of the point $(\L,\pi)\in 
H_{g,N}(1,\dots,1)$; 
$\B$ is the matrix of $b$-periods on $\L$;
$\tau$ is the isomonodromic  tau-function of Frobenius structure on $H_{g,N}(1,\dots,1)$
given by (\ref{tauint}).
\end{theorem}
The formula (\ref{detlap}) can be considered as a natural generalisation of the Ray-Singer formula
for the determinant of Laplacian on the torus with flat metric and periods $1$ and $\sigma$ \cite{Ray}:
\be
{\rm det}\Delta = C |\Im \sigma|^2 |\eta(\sigma)|^4
\la{detlap1}\ee
where $\eta$ is the Dedekind eta-function. The important feature of (\ref{detlap1}) is that the
function 
$$\f{{\rm det}\,\Delta}{{\{\Im\sigma\}\{Area(\L)\}}}$$
 is represented as the modulus square of a
holomorphic function on the moduli space. This is not the case for the higher genus formula (\ref{detlap}) due
to the presence of the factor $e^{{\mathbb S}_F}$, which does not admit the holomorphic factorization,
since the second order holomorphic-antiholomorphic derivatives of the logarithm of this function
are non-trivial \cite{Fay92,ZT}.

Actually, more natural higher genus analog of the Ray-Singer formula (\ref{detlap1}) is given by the
determinant of Laplacian computed in Strebel metrics (flat metrics with conic singularities), 
which are given by the modulus of holomorphic quadratic differential (or, in particular, by the modulus square of a holomorphic Abelian differential) \cite{Strebel}.

\subsection{Riemann-Hilbert problems with quasi-permutation monodromies 
and isomonodromic tau-function}

The Riemann-Hilbert problem of construction of $GL(N)$-valued function 
on the universal covering of punctured
Riemann sphere $\CP1\setminus\{\l_1,\dots,\l_M\}$ with prescribed 
monodromy representation in general case
(for an arbitrary representation) can not be solved in terms of known 
special functions. For an arbitrary
quasi-permutation monodromy group (i.e. such that each monodromy matrix 
has exactly 
one non-vanishing entry in each of its columns and each of its raws) 
the RH problem was solved in \cite{Kor03} 
outside of a divisor in the space of monodromy data (the so-called 
Malgrange divisor, or the divisor of zeros of 
the Jimbo-Miwa tau-function) following previous works 
\cite{KitKor,IKDZ}, where the $2\times 2$ case was solved.
One usually requires  the solution $\Psi$ of the Riemann-Hilbert 
problem to be normalized to the unit matrix at some point
$\l_0\in \CP1$, which does not coincide with singularities $\{\l_m\}$;
we shall denote such normalized solution by $\Psi(\l,\l_0)$.

\begin{theorem}\cite{Kor03}
Let the set of the monodromy data lie outside of   the Malgrange 
divisor.
Then the solution $\Psi(\l,\l_0)$ of an arbitrary Riemann-Hilbert
problem with 
quasi-permutation monodromy representation
is given by the analytical continuation on universal covering of the 
punctured sphere of the
following expression defined in a neighbourhood of the normalization 
point (all objects in this formula 
 correspond to the $N$-sheeted branched covering $\L$,
associated with the quasi-permutation monodromy representation):
\be
\Psi_{kj}(\l_0,\l)=\frac{\l-\l_0}{\sqrt{d\l d\l_0}}
\f{\Th\left[^\pb_\qb\right]\left(\Acal(\l^{(j)})-\Acal(\l_0^{(k)})+\O\right)}
{\Th\left[^\pb_\qb\right](\O)E(\l^{(j)},\l_0^{(k)})}\prod_{m=1}^M 
\prod_{l=1}^N \left[\f{E(\l^{(j)},\l_m^{(l)})}
{E(\l_0^{(k)},\l_m^{(l)})}\right]^{r_m^{(l)}}
\la{psinew}\ee
where 
\be
\O := \sum_{m=1}^M\sum_{j=1}^N r_m^{(j)} \Acal(\l_m^{(j)})\;;
\la{Om}\ee
$\l^{(k)}$ denotes the point of $\L$ which belongs to the $k$th sheet 
and has projection $\l$ on $\CP1$;
$\pb,\qb\in\C^g$ are constant vectors; $r_m^{(k)}$ are constants 
assigned to all points from $\pi^{-1}(\l_m)$
(if two points from $\pi^{-1}(\l_m)$ coincide, the constants  
$r_m^{(k)}$ are assumed to coincide, too).
The logarithms of the  matrix elements of  monodromy matrices are 
linear functions of the constants 
$\pb,\qb$ and  $r_m^{(k)}$. 
The Malgrange divisor is defined by the equation 
$\Th\left[^\pb_\qb\right](\O)=0$.
\end{theorem}
If the elements of monodromy matrices (or, equivalently, the constants 
$\pb,\qb$ and  $r_m^{(k)}$) are 
independent of positions of singularities $\{\l_m\}$, function $\Psi$ 
defines a solution of the Schlesinger system, together
with isomonodromic tau-function of Jimbo-Miwa \cite{JimMiw},
defined as follows:
\be
\f{\p}{\p\l_m}\log\tau_{1} = \f{1}{2}{\rm 
res}|_{\l=\l_m}\tr\left(\Psi_\l\Psi^{-1}\right)^2
\ee

In \cite{Kor03} it was proved the following

\begin{theorem}
The Jimbo-Miwa tau-function corresponding to solution (\ref{psinew})  
of the 
Riemann-Hilbert problem, is given by the following formula:
\begin{equation}
\tau_1=
\tau^{-1/2}\prod_{m,l=1}^M 
(\l_m-\l_l)^{r_{ml}}\Th\left[^\pb_\qb\right]\left(\O|\B\right)
\la{tauJM}
\end{equation}
where $\tau$ is the  tau-function defined by (\ref{deftau});
$$
r_{mn}=\sum_{k=1}^N r_m^{(k)} r_n^{(k)}\;.
$$
 \end{theorem}
This theorem, together with expression (\ref{tauint}) derived in this 
paper, gives the 
explicit formula for Jimbo-Miwa tau-function corresponding to general 
Riemann-Hilbert problem with quasi-permutation monodromies.
For
monodromy groups corresponding to hyperelliptic curves this tau-function 
was found in \cite{KitKor}; for
$Z_N$ curves with $N>2$  it was computed in \cite{GraEnol}.

We notice that the monodromy groups corresponding to fuchsian
Riemann-Hilbert problems of Hurwitz Frobenius manifolds, are not known
explicitly, in contrast to monodromy groups corresponding to solutions (\ref{psinew}). Therefore, one of the natural next problems is
to find the monodromy group and the solution of the Riemann-Hilbert problem which correspond 
to the tau-function (\ref{tauint}).

{\bf Acknowledgements}  The main results of this paper were obtained
during the stay of the authors at 
 Max-Planck-Institut f\"ur Mathematik in Bonn; we thank 
the institute  for warm hospitality and excellent working conditions.
This work was also partially supported by NSERC, NATEQ and Alexander von Humboldt Stiftung.

Department of Mathematics and Statistics, Concordia 
University 
\newline
7141 Sherbrooke West, Montreal H4B 1R6, Quebec,  Canada
\newline
e-mails: alexey@mathstat.concordia.ca; korotkin@mathstat.concordia.ca

\end{document}